\documentclass[showpacs,prd,aps,amsmath,amssymb,twocolumn,nofootinbib]{revtex4-1}
\usepackage{graphicx}
\usepackage{hyperref}

\newcommand{\beq}[1]{\begin{equation}\label{#1}} 
\newcommand{\eeq}{\end{equation}}
\newcommand{\beqa}{\begin{eqnarray}}
\newcommand{\eeqa}{\end{eqnarray}}
\newcommand{\apjl}{Astrophys.J.Lett.}
\renewcommand{\b}{\bar}
\newcommand{\cmsgev}{\ensuremath{\textnormal{cm}^3/\textnormal{s}/\textnormal{GeV}}}
\newcommand{\comment}[1]{}

\begin{document}

\title{Cosmic Microwave Background Constraints on Dark Matter Models of
the Galactic Center 511 keV Signal}

\author{Andrew R.\ Frey}\email[email: ]{a.frey@uwinnipeg.ca} 
\author{Nicholas B.\ Reid}
\affiliation{Department of Physics and Winnipeg Institute for Theoretical 
Physics, University of Winnipeg, Winnipeg, Manitoba, Canada R3B 2E9}

\begin{abstract}
The high positron production rate required to explain the flux of 511 keV
gamma rays from the galactic center has inspired many models in which dark
matter creates positrons.  These models include the annihilation of light
dark matter and scattering of dark matter with excited states
(exciting dark matter).  We show that existing cosmic microwave background
data robustly constrains such models when the annihilation or scattering
cross section is not velocity suppressed depending on the model of the
galactic dark matter halo.  Upcoming data from the Planck mission can exclude 
the fiducial \textit{Via Lactea II} halo model, which also provides a good
fit to the 511 keV morphology.  We additionally find combined constraints on
exciting dark matter scattering and annihilation and update constraints on
the lifetimes of dark matter excited states.  Finally, we apply
constraints to models of dark matter decay in which produced positrons fall
into the galactic center and produce the 511 keV signal on their annihilation,
demonstrating that most of the parameter space of interest is ruled out.
\end{abstract}
\maketitle

\section{Introduction}

A narrow line of 511 keV gamma rays from the galactic center has been
observed since the 1970s \cite{jhh,lms}, indicating annihilation of 
$\gtrsim 10^{43}$ electron-positron ($e^\pm$) pairs per second.  
This emission has been studied extensively by INTEGRAL/SPI
since 2002 \cite{Knodlseder:2003sv,Jean:2003ci,Knodlseder:2005yq,
Churazov:2004as,Jean:2005af,Weidenspointner:2008zz,bouchet:2008,
Bouchet:2010dj}.  The spectrum indicates that $\sim 97\%$ of the $e^+$ 
annihilate through positronium formation, which limits the $e^+$ injection
energy to a few MeV or less, and a more detailed analysis finds that 
a significant majority of the annihilation occurs in warm neutral and
ionized phases of the ISM.  Morphologically, the emission has two
components, from the galactic center bulge and from the galactic disk.
Depending on the model used, the bulge/disk ratio for $e^+$ annihilation
rates is $\sim 1.5-6$.  

Over the years, many models for the astrophysical production of mildly
relativistic $e^+$ have been proposed, although standard mechanisms have
difficulty reproducing the observed morphology of the emission, particularly
the large bulge/disk ratio; \cite{Prantzos:2010wi} gives a thorough
review of not only the emission itself but also many $e^+$ production
mechanisms.  One which has received a great deal of attention is $\beta^+$
decay of radionuclides produced in supernovae and heavy stars, which
reproduces the disk emission well.  \cite{Higdon:2007fu,Lingenfelter:2009kx}
have argued that, in a particular model of $e^+$ transport, radionuclide
decay can also source the bulge emission.  On the other hand, 
as the review \cite{Prantzos:2010wi} notes,
the model of \cite{Higdon:2007fu,Lingenfelter:2009kx} relies on a number
of apparently arbitrary 
assumptions about the interstellar medium and magnetic field
(and its turbulence) in the bulge.  Recently, \cite{Martin:2012hv}
found that low-energy positrons annihilate close to their production
sites in a range of propagation models, so, in contradiction to 
\cite{Higdon:2007fu,Lingenfelter:2009kx}, they find that radionuclides
(produced mainly in the disk) cannot source the bulge emission.  In any
event, it is clearly premature to claim that standard astrophysical
mechanisms can produce the galactic bulge positrons.

Due to the lack of consensus on candidate astrophysical sources for 
the galactic bulge emission, more exotic mechanisms have been of great
interest in the past decade, including a number involving dark matter (DM).
Roughly, DM models for the 511 keV excess may produce $e^\pm$ pairs
through the decay or scattering of DM particles.  There are two cases of 
DM decay that produce $e^+$ with only mildly relativistic injection energies:
first, the DM itself can have an MeV-scale mass 
\cite{Picciotto:2004rp,Hooper:2004qf}, or the DM can have several
states with the decay through an MeV-scale mass gap producing an $e^\pm$ pair
\cite{Cline:2010kv}.
In either case, assuming small $e^+$ propagation after production,
the 511 keV flux is proportional to the integral of $\rho_{DM}$ along the
line of sight.  
As the DM density is expected to increase toward the galactic center, 
all these models naturally have a large bulge/disk ratio for $e^+$ production.
However, \cite{Vincent:2012an,Ascasibar:2005rw}
showed that the INTEGRAL signal is 
more highly peaked toward the center of the galaxy than realistic
DM density profiles (even the most cuspy found in simulations), 
essentially ruling out decay of DM at either mass
range as the source of the galactic bulge $e^+$ excess.
Recently, \cite{Boubekeur:2012eq} proposed a novel model in which a small
fraction of DM decays, producing $e^+$ outside the galaxy; these $e^+$ fall
into the galaxy and annihilate (preferentially in the galactic center,
as claimed in \cite{Boubekeur:2012eq}).

Alternately, DM can produce $e^\pm$ pairs through scattering processes.
In the simplest such models, the DM has an MeV-scale mass, and the scattering
process is direct annihilation to an $e^\pm$ pair.  These models and
additional signatures have been studied extensively in 
\cite{Boehm:2003bt,Hooper:2003sh,Fayet:2004bw,Casse:2004gw,Cordier:2004hf,
Ascasibar:2005rw,Zhang:2006fr,Mapelli:2006ej,Huh:2007zw,
deNiverville:2011it,Ho:2012ug,Ho:2012br}.  Alternately, more massive DM
with several states and MeV-scale mass gaps can scatter into an unstable
excited state, which decays into the ground state via emission of an $e^\pm$
pair, as discussed in \cite{Finkbeiner:2007kk,Pospelov:2007xh,
Finkbeiner:2008gw,ArkaniHamed:2008qn,Chen:2009dm,Finkbeiner:2009mi,
Batell:2009vb,Chen:2009ab,Chen:2009av,Cline:2010kv,Morris:2011dj,Cline:2012yx,
Bai:2012yq}.  (These are exciting dark matter, or XDM, models.)
In these scattering models, the gamma ray flux follows 
the line of sight integral of $\rho_{DM}^2$, so reasonable halo models
can produce the observed signal; in fact, \cite{Vincent:2012an} has found 
that the halo model derived in the \textit{Via Lactea II} simulation
\cite{Diemand:2008in,Kuhlen:2009kx} has a maximum likelihood ratio very 
close to the peak value.  

The lack of a strong astrophysical candidate for the galactic bulge
emission as well as the surprising (and striking) agreement between the 
emission morphology and simulated DM halos motivates us to search for
other potential signals of DM scattering models either as circumstantial
evidence in favor of these models or as constraints on them in the case
of non-observation.  For example, the XDM models discussed in 
\cite{ArkaniHamed:2008qn,Chen:2009ab,Chen:2009av,Cline:2010kv,Cline:2012yx} 
naturally include
a several hundred MeV gauge boson with weak coupling to electric charge,
which could be discovered at fixed target experiments such as the
Mainz Microtron \cite{Merkel:2011ze} or APEX \cite{Abrahamyan:2011gv}.
In this paper, we place constraints on XDM and light annihilating DM 
explanations based on limits from the cosmic microwave background (CMB)
anisotropy spectrum; as we will review, energy injected into the Standard 
Model (SM) plasma around and after the era of recombination is tightly 
constrained by large $l$ CMB anisotropies.  As long as the XDM scattering
or light DM annihilation cross section is not suppressed at low velocities,
we find that these models for the 511 keV emission will be constrained
by forthcoming results from the Planck satellite \cite{Ade:2011ah}, 
specifically ruling out the preferred DM halo parameters from
\textit{Via Lactea II} (although there is sufficient uncertainty in
both halo and DM model parameters to find a small allowed region 
consistent with \textit{Via Lactea II}).

The plan of this paper is as follows.  In the next section, 
we systematically review DM models for the 
production of mildly relativistic $e^+$ in the galactic center.  
In section \ref{xdmconstraints}, 
we present constraints on $e^\pm$ creation during (and after) recombination 
from CMB anisotropies and discuss how they constrain XDM and
annihilating MeV-mass DM models.  Next, since XDM models contain decaying
excited states of DM, we discuss how CMB observations provide constraints 
on those lifetimes, updating the results of \cite{Finkbeiner:2008gw},
and also place constraints on the recent model of \cite{Boubekeur:2012eq}.
We conclude with a brief discussion of our results.

\section{Dark Matter Models for the 511 keV Emission}\label{xdmmodels}

As discussed above, the morphology of the 511 keV emission rules out models
of DM decay as an explanation for positron production in the galactic
center, so we will focus on models in which DM scattering
processes produce $e^\pm$ pairs.  The rate of positron production in
such models is
\beq{eplusrate}
R=\eta s \bar Y^2 \int_{\textnormal{bulge}} d^3\vec x\, \left(
\frac{\rho_{DM}(\vec x)}{M}\right)^2 \langle\sigma v_{rel}\rangle (\vec x)
\ ,\eeq
where $M$ is the DM mass, $s$ is $1/2$ for real/Majorana DM 
and $1/4$ for complex/Dirac DM, $\eta$ is the number of $e^+$ produced per
scattering event, and $\bar Y=Y/Y_{DM}$ is the relative abundance of the
active DM state.  $\sigma$ is the appropriate scattering cross section
with relative velocity $v_{rel}$, and $\langle\cdots\rangle$ is the 
average over the DM velocity distribution at position $\vec x$.
We further write $\langle\sigma v_{rel}\rangle =\overline{\sigma v}
\langle F(v_{rel})\rangle$, where $\overline{\sigma v}$ is the cross section
at some representative relative velocity and $F$ is a dimensionless
model-dependent function of the relative velocity.  The $e^+$ production
rate is $R=1.1\times 10^{43}$ s$^{-1}$ \cite{Bouchet:2010dj}.

We follow \cite{Cline:2012yx} in writing
\beq{ratezeta}
R=4\pi\eta s\b Y^2 \rho_\odot^2\, \frac{\overline{\sigma v}}{M^2}\,\zeta\,
(\textnormal{kpc})^3\ ,\eeq
where $\rho_\odot$ is the DM density in the solar neighborhood, which we 
take to be $\rho_\odot=0.4$ GeV/cm$^3$ \cite{Salucci:2010qr}.  $\zeta$ is
\beq{zetadef}
\zeta=\textnormal{kpc}^{-3}\int_0^{r_c} dr\, r^2 (\rho_{DM}(r)/\rho_\odot)^2
\langle F(v_{rel})\rangle\ ,
\eeq
where we take the bulge radius $r_c=1.5$ kpc, corresponding to the width
of the INTEGRAL signal.  
We assume that the DM density follows an Einasto profile
\beq{einasto}\rho_{DM}(r)/\rho_\odot=\exp\left[
-(2/\alpha)\left( (r/r_s)^\alpha - (r_\odot/r_s)^\alpha \right)\right]\eeq
with the sun located at $r_\odot=8.5$ kpc and
$\alpha$ and $r_s$ as free parameters.  The \textit{Via Lactea II}
simulation is fit by $\alpha=0.17$, $r_s=25.7$ kpc, which also lies in the
best fit region of \cite{Vincent:2012an} for the 511 keV signal.  Smaller
values of $\alpha$ and $r_s$ lead to a more cuspy central halo.
We assume that DM velocities follow a Maxwell distribution with dispersion
$v_0(r)$ cut off at the escape velocity $v_{esc}(r)$ satisfying
\beqa
v_0(r)^3 &\propto& r^{1.64}\,\rho(r)\ ,\ \ v_0(r_\odot)=220-230 \textnormal{km/s}\\
v_{esc}(r)^2 &=& 2 v_0(r)^2\left[2.39 + \ln(10\ \textnormal{kpc}/r)\right]
\eeqa
as in \cite{Cline:2010kv} (the choice of escape velocity follows
\cite{Cirelli:2010nh}, and the velocity dispersion is suggested by 
simulations including baryonic contraction \cite{2010MNRAS.406..922T}).

We will consider three models in which DM scattering processes produce
mildly relativistic positrons.  The first is annihilation of MeV-scale
DM, and the following two are endothermic and exothermic XDM respectively.
Finally, we will review a recent model by \cite{Boubekeur:2012eq} in which
decays of a metastable DM component creates $e^+$ at late time which
fall into the galactic bulge.

\begin{figure}[t]
\includegraphics[scale=0.65]{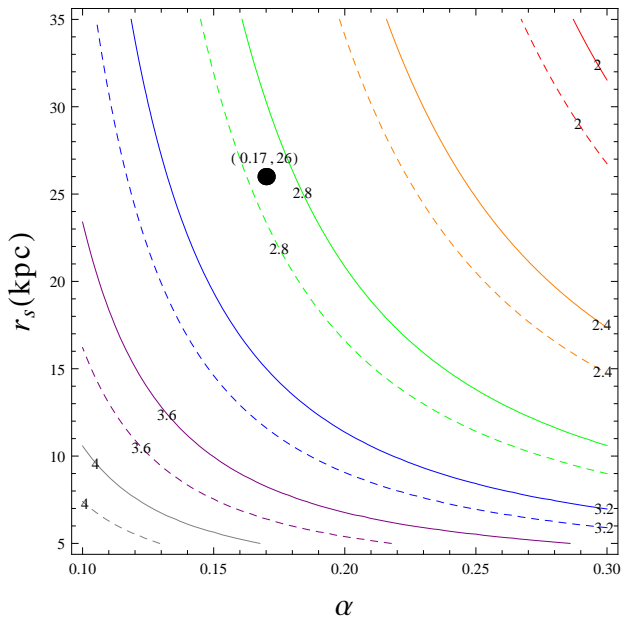}
\caption{\label{f:zetacontour} Contours of $\log\zeta_{ann}$ (dashed) and 
$\log\zeta_\downarrow$ (solid) as a function of the Einasto profile 
parameters $\alpha$ and $r_s$.  The dot represents the \textit{Via Lactea II}
parameters. $\zeta_\downarrow$ is calculated for $v_t=10^{-3}$.}
\end{figure}

\subsection{Annihilation of Light Dark Matter}\label{mevann}

As first proposed by \cite{Boehm:2003bt}, 
DM particles of mass $2m_e<M\lesssim 10$ MeV have a sufficient $e^+$
production rate if they annihilate into $e^\pm$ pairs with the appropriate 
cross section, $\overline{\sigma v}\sim 10^{-31} (M/\textnormal{MeV})^2$
cm$^3$/s. As this cross section is several orders of magnitude smaller than 
required for the correct thermal relic abundance, either the annihilation
is $p$-wave dominated or annihilation to $e^\pm$ is subdominant to another
channel, as in \cite{Huh:2007zw}.  Morphological studies of the 511 keV
signal using a similar profile for the velocity dispersion\footnote{Our
velocity dispersion decreases slightly less rapidly as $r\to 0$ for a 
fixed density profile.} 
disfavors $p$-wave annihilation \cite{Ascasibar:2005rw}, so we consider
only the $s$-wave cross section.  In this case, $F(v_{rel})=1$ and $\zeta$
is simply the integral of $\rho_{DM}^2$.  Figure \ref{f:zetacontour} shows 
contours of $\log\zeta$ (henceforth denoted as $\zeta_{ann}$ for this model) 
as dashed lines as a function of $\alpha$ and $r_s$.  Additional 
signatures of these models have been considered in 
\cite{Hooper:2003sh,Casse:2004gw,Cordier:2004hf,Huh:2007zw,deNiverville:2011it,
Ho:2012ug,Ho:2012br}.  Finally, \cite{Zhang:2006fr,Mapelli:2006ej} placed
constraints on MeV DM annihilation to $e^\pm$ based on the CMB; our results
extend and update those works.

\comment{
A number of authors have considered additional predictions of this model.
\cite{Hooper:2003sh,Casse:2004gw,Cordier:2004hf} studied the predicted 
511 keV emission from nearby dwarf spheroidal galaxies, which is consistent
with observed limits (these predictions should be similar for XDM models).
\cite{deNiverville:2011it} demonstrated that models like those of 
\cite{Huh:2007zw} with additional light states could be detected in 
neutrino oscillation experiments.  Recently, \cite{Ho:2012ug,Ho:2012br} 
demonstrated
that DM with MeV masses annihilating preferentially into $e^\pm$ as opposed
to $\nu$ can alter Big Bang nucleosynthesis, providing an alternate set
of constraints.  Finally, \cite{Zhang:2006fr,Mapelli:2006ej} placed
constraints on MeV DM annihilation to $e^\pm$ based on the CMB; our results
extend and update those works.
}

\subsection{Endothermic Exciting Dark Matter}\label{endo}

XDM as first envisioned by 
\cite{Finkbeiner:2007kk,Pospelov:2007xh,ArkaniHamed:2008qn}
consist of a DM ground state (state \#1) and an unstable excited state 
(state \#2) with a mass splitting $\delta M_{12}>2m_e$.  
DM collisions above a threshold velocity $v_t$ populate the excited state,
which then decays to the ground state by releasing an $e^\pm$ pair.
With $M\sim$ TeV and $\delta M_{12}\gtrsim 2m_e$, $v_t$ is approximately the 
velocity dispersion of DM in our galaxy.  In order to reduce $v_t$ or to 
accommodate lower DM masses (to $\sim 5-10$ GeV), consider that
a significant fraction of DM remains in a metastable excited state (state \#3),
and upscattering of that state into the unstable state through a smaller
mass gap $\delta M_{32}$, allowing a sufficient $e^+$ production rate 
\cite{Chen:2009dm,Finkbeiner:2009mi,Chen:2009ab,Chen:2009av,Cline:2010kv,
Morris:2011dj}.  Direct detection of XDM has been considered in
\cite{Finkbeiner:2009mi,Batell:2009vb,Chen:2009ab,Cline:2010kv}.

In these models, the fraction $\b Y$ of DM in the 
metastable excited state is determined by freeze out of DM-DM scattering,
as detailed in \cite{Cline:2010kv}.  
A potentially significant fraction of DM can be
initially (after chemical and kinetic freeze out) in the unstable excited
state.  \cite{Finkbeiner:2008gw} has provided limits on the lifetime 
of this decay based on CMB measurements; our results in section 
\ref{decayingconstraints} will update those constraints.

In these models, $F(v_{rel})=\sqrt{v_{rel}^2/v_t^2 -1}\,
\Theta(v_{rel}-v_t)$,
including both a phase space suppression and the kinematic threshold.
As the kinetic energy of DM at recombination and later
(before structure formation and virialization) is much too small to 
allow this upscattering, 
endothermic XDM is not subject
to the CMB constraints discussed in section \ref{xdmconstraints} below.

\subsection{Exothermic Exciting Dark Matter}\label{exo}

Another possibility is that the metastable DM excited state \#3 scatters
exothermically into the unstable excited state \#2 \cite{Cline:2010kv}; 
in fact, this possibility is preferred when the DM is charged under an 
abelian gauge group \cite{Cline:2012yx}.  This downscattering process
is not kinematically suppressed, so it can produce $e^\pm$ pairs throughout
recombination. In this case, $F(v_{rel})=\sqrt{v_{rel}^2/v_t^2+1}$.\footnote{There
is a weak additional velocity dependence; details for scattering by gauge
boson exchange are given in \cite{Cline:2010kv}.}  Very recently,
\cite{Bai:2012yq} proposed an exothermic XDM model that also produces a 
gamma ray line at the DM mass, simultaneously explaining a line at 130-135
GeV in galactic center observations of the Fermi satellite 
\cite{Bringmann:2012vr,Weniger:2012tx,Tempel:2012ey,Su:2012ft,fermi135} 
(see \cite{Bringmann:2012ez} for a review).

In exothermic XDM, we define the velocity $v_t$ as the threshold velocity
for the inverse upscattering process (or equivalently, as the velocity
imparted to the less massive states for downscattering at rest).
This is given by 
$v_t = 2\sqrt{2\delta M_{23}/M}$ when both DM particles excite or
de-excite in the scattering.\footnote{Note a factor of 2 redefinition of 
$v_t$ compared to \cite{Cline:2010kv}.}  For example, a mass of $M=10$ GeV
and $v_t=10^{-3}$ correspond to a mass splitting of $1.25$ keV between
the two excited states of DM.
The solid contours in figure \ref{f:zetacontour} show contours of $\log\zeta$
(henceforth $\zeta_\downarrow$ for exothermic XDM) as a function of the 
Einasto parameters $\alpha$ and $r_s$ for $v_t=10^{-3}$.  Compared to
$\zeta_{ann}$ at a fixed $\alpha,r_s$, $\zeta_\downarrow$ is enhanced, and 
the enhancement increases as $v_t$ decreases.  As a result, contours of
$\log\zeta_\downarrow$ shift to the right as $v_t$ decreases and approach the
contours of $\log\zeta_{ann}$ as $v_t\to\infty$.  Since $v_t\lesssim 10^{-3}$ 
typically requires additional fine-tuning, we consider the contours of
figure \ref{f:zetacontour} to give an estimate of the uncertainty in our 
modeling.  In addition, constraints placed on models with $v_t=10^{-3}$ will
be weaker than models with larger $v_t$, so we consider it to be a conservative
choice.

The other phenomenology of exothermic XDM is similar to that of endothermic
XDM.  In some cases, the metastable state can decay to the unstable state
by emission of a single X-ray photon; this signal can be near observable
levels \cite{Cline:2010kv}.

\subsection{Infalling Positrons from Decaying Dark Matter}\label{infall}

Recently, \cite{Boubekeur:2012eq} proposed a new model of decaying
DM which could provide a sufficient number of $e^+$ to explain the
511 keV signal.  In their model, DM has two states, a stable ground state
and a metastable excited state with mass splitting $\delta M\lesssim$ GeV.
The excited state can decay to the ground state by emitting $e^+$, and
the constraint on the mass splitting prevents antiproton production
(although a more general model may be acceptable from that point of view).
The $e^+$ cool to nonrelativistic energies via scattering processes; 
some $e^+$ in large orbits are only entering the galaxy and annihilating 
in the present day.  Assuming the ground and excited states are similar
in mass, \cite{Boubekeur:2012eq} found that $\bar Y_{rec}/M\sim 5\times 10^{-8}$
GeV$^{-1}$ yields a sufficiently strong gamma ray signal (as above,
$\bar Y_{rec}$ is the fraction of DM in the excited state at recombination).
This model is free from the morphological constraints of 
\cite{Vincent:2012an,Ascasibar:2005rw} because the DM decays occur outside
the galaxy, but no detailed morphological study has yet been performed.
In section \ref{decayingconstraints}, we will find constraints on this class
of models independent of details of the decay.  

\section{Constraints on Scattering and Annihilating Dark Matter}
\label{xdmconstraints}

Any energy injected into the SM inter-galactic medium (IGM) 
from a hidden sector (as by weakly interacting massive particle (WIMP)
annihilation)
after matter-radiation equality modifies the 
recombination history of the universe, which increases residual ionization.
That increases the optical depth of the IGM 
and thickness of the last scattering surface, 
which slightly alters the CMB anisotropy spectrum, suppressing high
multipole moments.  
As a result, there is a wide literature
deriving constraints on excess energy injection based on CMB measurements,
particularly due to DM decay or annihilation.
For example, see \cite{Chen:2003gz,Pierpaoli:2003rz,Padmanabhan:2005es,
Zhang:2006fr,Mapelli:2006ej,Zhang:2007zzh,Galli:2009zc,Cirelli:2009bb,
Slatyer:2009yq,Hutsi:2011vx,Finkbeiner:2011dx,ArmendarizPicon:2012mu,
Yeung:2012ya,Evoli:2012qh,Farhang:2012jz}. 

The physically important quantity is $dE/dVdt$, 
the rate of energy deposition to the SM IGM per volume.  
It is common to separate astrophysical factors from
model-dependent particle physics; for a DM scattering or annihilation
process, 
\beq{pann1}
\frac{dE}{dVdt}= \rho_c^2 \Omega_{DM}^2 (1+z)^6 (p_{ann}+p_{scatt})\ .
\eeq
We have defined 
\beq{pann2}
p_{ann} = 2 f s \frac{\langle\sigma_{ann} v_{rel}\rangle}{M}\eeq
for annihilation cross section $\sigma_{ann}$ (into SM particles) and
\beq{pscatt1}
p_{scatt} = \bar Y^2 fs \eta\delta M_{12}
\frac{\langle\sigma_\downarrow v_{rel}\rangle}{M^2} \eeq
where $\sigma_\downarrow$ is the downscattering cross section in the case
of exothermic dark matter.  Here $f$ is an energy deposition efficiency
discussed further below, and other
variables are defined as in section \ref{xdmmodels} above.  
Constraints on the modification of 
the CMB anisotropies currently exclude some
annihilation channels for thermal WIMPs of mass $\lesssim 10$ GeV
and are weakened in asymmetric
DM models \cite{Lin:2011gj,Slatyer:2012yq} or if the DM dominantly
annihilates into hidden sector particles \cite{Cline:2011uu}.

The deposition efficiency $f$ depends on the specific particle physics model
of DM through the branching ratios to different decay/annihilation products.
Since the IGM is transparent to photons of certain energies 
at certain redshifts, $f$ depends on redshift $z$, as well, but 
\cite{Galli:2011rz} noted
that the full redshift dependence is well approximated (within 15\%) 
for DM annihilation by
taking $f$ to be constant at the value for $z=600$.
For moderately relativistic $e^\pm$, \cite{Slatyer:2012yq} finds 
$f(z=600)>0.9$, and we adopt that constant value (the error associated with
the approximation is smaller than the allowed range in $M,\delta M_{12}$ for 
annihilating light DM or XDM models).

We will consider several experimental constraints.  First, 
\cite{Galli:2011rz} finds a 95\% confidence constraint of 
$p_{ann}<2.42\times 10^{-27}$ \cmsgev\ from WMAP7 data \cite{Larson:2010gs}, 
$p_{ann}<2.09\times 10^{-27}$
\cmsgev\ from WMAP7 and ACT \cite{Fowler:2010cy}, and a forecast 
constraint of $p_{ann}<3.03\times 10^{-28}$ \cmsgev\ from Planck.
Finally, \cite{Giesen:2012rp} find $p_{ann}<7.86\times 10^{-28}$ \cmsgev\
using WMAP7 and SPT \cite{Keisler:2011aw} data at 95\% confidence. 
These apply equally to $p_{scatt}$.

\subsection{Conservative Constraints}

\begin{figure}
\includegraphics[scale=0.65]{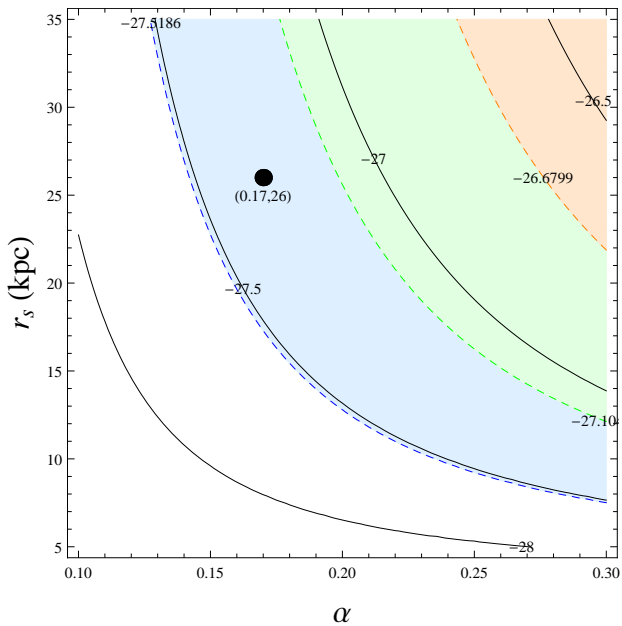}
\caption{\label{f:mevconstraints} Contours of $\log p_{ann}$ (in \cmsgev)
for $s$-wave annihilating light DM for DM mass $fM=0.75$ MeV with 95\%
confidence constraint surfaces (dashed contours).  Regions to the right of
the constraint contours are ruled out.  The rightmost constraint 
(orange) is ruled out by
WMAP7+ACT \cite{Galli:2011rz} and the middle constraint (green)
by WMAP7+SPT \cite{Giesen:2012rp}.  The left constraint (blue)
will be ruled out by forecast constraints from Planck \cite{Galli:2011rz}.
The dot indicates the \textit{Via Lactea II} parameters.}
\end{figure}

\begin{figure}
\includegraphics[scale=0.65]{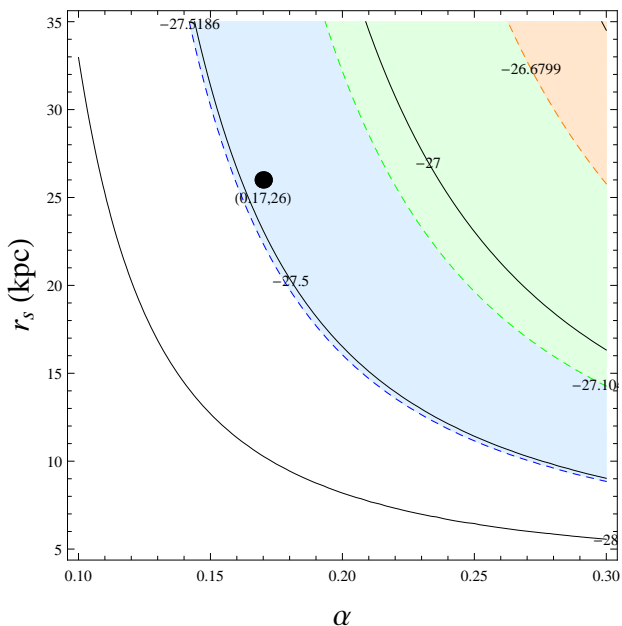}
\caption{\label{f:xdmconstraints} Contours of $\log p_{scatt}$ (in \cmsgev)
for XDM with $v_t=10^{-3}$ and $f\delta M_{12}=1.5$ MeV with 95\%
confidence constraint surfaces (dashed contours).  Constraints are 
labeled as in figure \ref{f:mevconstraints}.  The dot indicates the 
\textit{Via Lactea II} parameters.}
\end{figure}

The astute reader will notice a similarity between equations (\ref{ratezeta})
and (\ref{pann2},\ref{pscatt1}), which allows us, assuming that DM scattering 
is responsible for the $e^+$ production in the galactic bulge,
to find $p_{ann}$ or $p_{scatt}$ as functions of the Einasto profile parameters
$\alpha$ and $r_s$.  The key point to understand is the relation between
$\langle\sigma_{ann,\downarrow} v_{rel}\rangle$ in $p_{ann}/p_{scatt}$, 
which is the thermal average
in the early universe, to $\overline{\sigma v}$.  For $s$-wave annihilation,
as appropriate for the MeV-mass annihilating DM of section \ref{mevann},
these quantities are equal.  For XDM models with DM mass $M\gtrsim$ GeV,
the DM velocity dispersion at recombination and later (but before structure 
formation) is considerably smaller than $v_t$.  Therefore, upscattering
in endothermic XDM models is kinematically forbidden as noted, and 
$\langle\sigma_{\downarrow} v_{rel}\rangle=\overline{\sigma v}$ for exothermic
XDM since $F(v_{rel}=0)=1$.  

As a result, we find that production of low-energy $e^+$ by exothermic
scattering processes minimally deposits 
\beq{pscatt2}
p_{scatt}= \left(R/4\pi \rho_\odot^2 \zeta_\downarrow \textnormal{kpc}^3\right)
f\delta M_{12}\ .\eeq
This includes annihilation of light DM if we replace 
$\zeta_\downarrow\to\zeta_{ann}$ and set $\delta M_{12}= 2M$.\footnote{Annihilation
of DM produces one $e^\pm$ pair per DM pair vs one $e^\pm$ pair per
excited DM state in XDM.}  As a result, 
constraints on $p_{ann}$/$p_{scatt}$ rule out regions of $\alpha,r_s$
parameter space for light DM annihilation or exothermic XDM, assuming
a fixed value of $\delta M_{12}$.  Note that equation (\ref{pscatt2}) is
independent of $M$ except through a relatively weak 
dependence on $v_t$, which is determined by the ratio $\delta M_{23}/M$.
We will assume that $f$ is constant in an on-the-spot approximation; a 
more detailed parameter search using the redshift and energy dependence of
$f$ may be warranted in the future.

Constraints on the allowed Einasto profile halo parameters are shown 
in figure \ref{f:mevconstraints} for annihilating light dark matter and
in figure \ref{f:xdmconstraints} for exothermic XDM.  For illustrative
purposes, we have chosen $2fM,f\delta M_{12}=1.5$ MeV; this is a representative
but conservative choice in that generated $e^+$ are mildly relativistic,
lower $e^+$ energies require tuning in the DM model, and the constraints
become more stringent for larger $e^+$ energies ($M$ or $\delta M_{12}$).
For reference, the \textit{Via Lactea II} halo model for annihilating light DM 
is already ruled out by the 95\% confidence constraints from WMAP7+SPT 
\cite{Giesen:2012rp} if $fM>1.2$ MeV, and it is always ruled out by the
forecast constraints from Planck given in \cite{Galli:2011rz} (unless
the efficiency factor $f$ is for some reason reduced below expected values).
As discussed earlier, we also choose $v_t=10^{-3}$ in figure 
\ref{f:xdmconstraints} as a conservative choice without fine tuning; for
this $v_t$, the \textit{Via Lactea II} model is already ruled out
by WMAP7+SPT if $\delta M_{12}>3.6$ MeV (assuming $f=0.9$) and would be free
of the forecast Planck constraints if $f\delta M_{12}<1.2$ MeV.  For larger
values of $v_t$, the constraints approach those shown in figure 
\ref{f:mevconstraints}.

Similar constraints for annihilating light DM have been discussed 
previously in \cite{Zhang:2006fr}, which were based on the NFW profile
selected by \cite{Ascasibar:2005rw}.  While this profile is slightly
cuspier in the innermost galaxy 
than the \textit{Via Lactea II} Einasto profile, the corresponding
value of $\zeta_{ann}$ is actually slightly less for the NFW profile assuming
that both profiles are normalized to $\rho_\odot$ at $r_\odot$.
Furthermore, \cite{Ascasibar:2005rw} took $\rho_\odot=0.3$ g/cm$^3$.
Making the appropriate conversion, \cite{Zhang:2006fr} constrains 
annihilating light DM to have $fM<1.7$ MeV.  As expected, the improvement
in the CMB data since \cite{Zhang:2006fr} has tightened the constraints.

Finally, we note that these constraints are conservative because they arise
from the same mechanism that produces $e^+$ in the galactic bulge
\textit{as long as the scattering is not velocity-suppressed} (or kinetically
forbidden at the time of recombination as in endothermic XDM).
As a result, upcoming results from Planck will be able to exclude 
annihilating light DM and exothermic XDM at the \textit{Via Lactea II}
halo parameters (assuming that the mass splitting $\delta M_{12}$
and threshold $v_t$ are not finely tuned in the case of XDM).  In fact,
the forecast Planck constraints exclude a significant fraction of the best-fit 
parameter space from \cite{Vincent:2012an}, though smaller values of $r_s$
are allowed for XDM as long as $\delta M_{12}$ is not too large.  
Aside from fine-tuning of
the particle physics, our results suggest one additional way these models
might evade CMB constraints.  More cuspy DM halo profiles (with larger values
of $\zeta$) will be unconstrained by CMB experiments, and propagation of
unstable DM excited states or the produced $e^+$ themselves can, in principle,
spread out the 511 keV gamma ray signal.  However, the similar morphology
of the 511 keV and 130-135 GeV lines (pointed out in \cite{Bai:2012yq}) 
makes this way out less palatable, assuming that the 130 GeV signal withstands
further scrutiny.

\subsection{Annihilation of XDM}

XDM can also annihilate completely, depositing additional energy to the
IGM, like a standard thermal WIMP.  In the exothermic case, the total 
energy deposition is given by $p_{ann}+p_{scatt}$ in equation (\ref{pann1}).
It is already possible to rule out thermal WIMPs
with mass in the range of 7 to 12 GeV, which has been of
interest with respect to possible signals at the DAMA \cite{Bernabei:2008yi},
CoGeNT \cite{Aalseth:2010vx,Aalseth:2011wp}, and CRESST \cite{Angloher:2011uu} 
direct detection experiments, using the WMAP7+SPT constraints discussed
above.  In the following, we will illustrate that production of low-energy
$e^\pm$ pairs provides extra sensitivity in the case of exothermic XDM 
in a model-dependent fashion.  We focus on the
exothermic XDM models of \cite{Cline:2010kv} as an example, working under
a few simplifying assumptions.

In these models, DM is a Majorana fermion triplet of a dark $SU(2)$ 
gauge group and annihilates into the dark gauge bosons, assuming that the 
dark Higgs bosons are heavier than the DM.  The dark gauge bosons all
mix kinetically with the photon, so they can decay into any lighter charged
SM particle.  Then, for light gauge bosons, the final annihilation products are
$e^\pm$, $\mu^\pm$, and $\pi^\pm$.  This allows us to find an average 
efficiency factor $f$ for the annihilation as the average of the 
``XDM electrons,'' ``XDM muons,'' and ``XDM pions'' values of 
\cite{Slatyer:2009yq} weighted by the gauge boson branching ratios as a 
function of the gauge boson mass.  For simplicity and specificity, 
we take all the gauge bosons to have the same mass of 500 MeV, which yields
$f=0.53$.

\begin{figure*}
\includegraphics[scale=0.4]{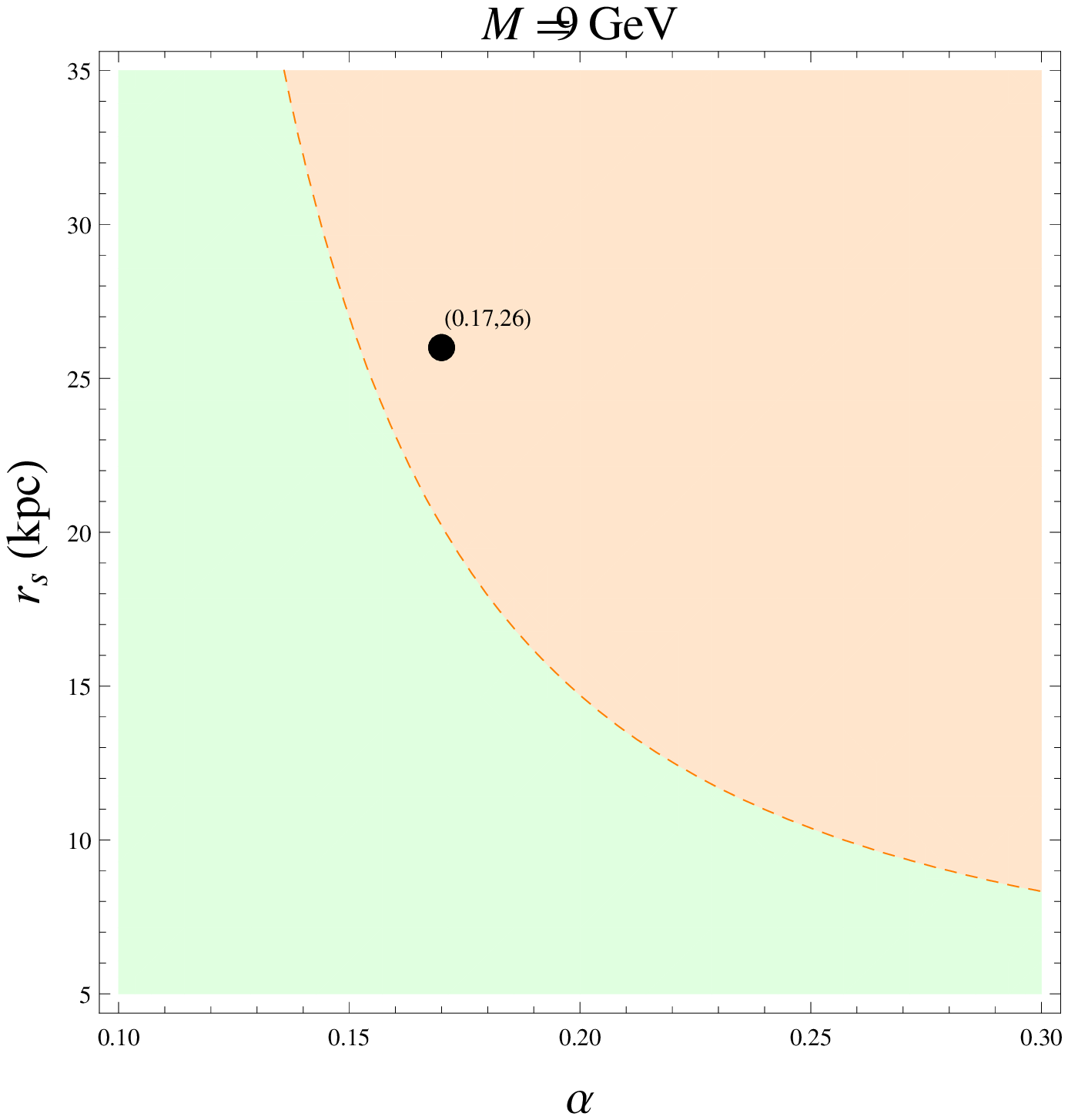}\ 
\includegraphics[scale=0.4]{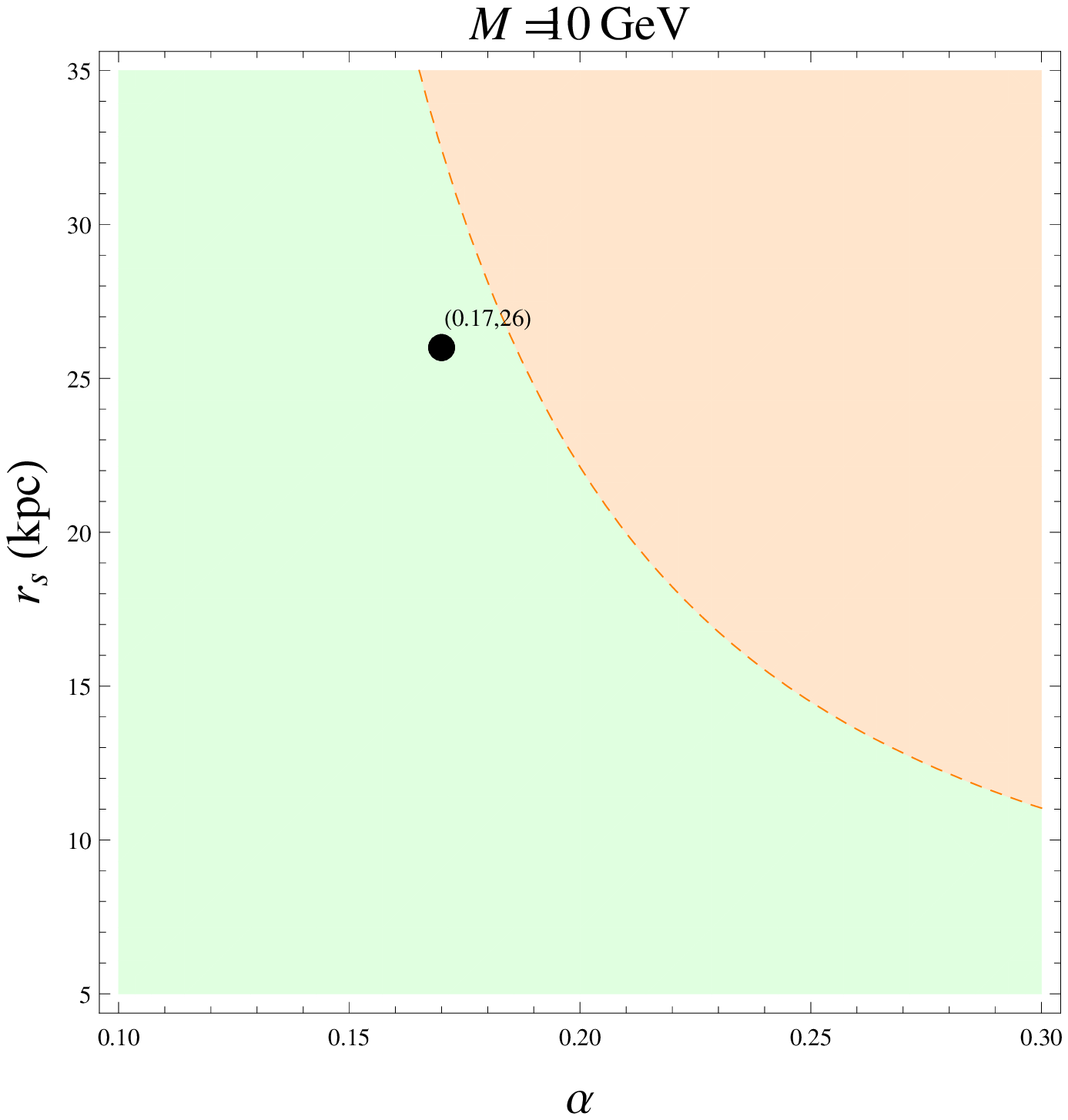}\ 
\includegraphics[scale=0.4]{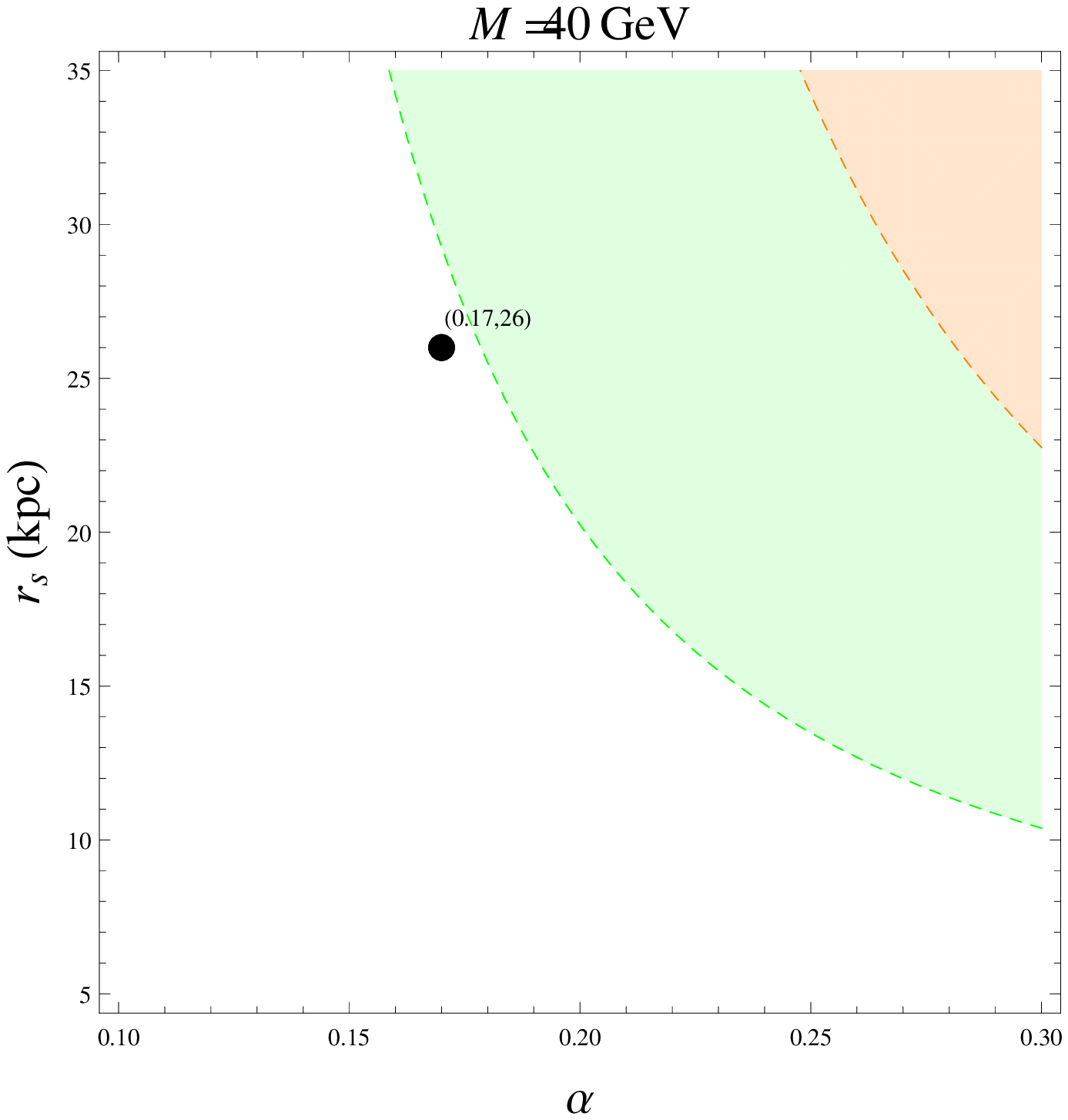}
\caption{\label{f:annihilating} Exclusion plots for a specific XDM model
including energy deposition from DM annihilation and scattering for
DM mass $M=$ 9 GeV (left panel), 10 GeV (center), and 40 GeV (right).  
Contours are of $p_{ann}+p_{scatt}$ at the WMAP7+ACT constraint (orange) and
the WMAP7+SPT constraint (green).  Excluded regions are shaded.}
\end{figure*}

In addition, the annihilation cross section at late times,
$\langle\sigma_{ann}v_{rel}\rangle$, differs from the canonical value
$3\times 10^{-26}$ cm$^3$/s for three reasons.  First, as discussed in
\cite{Steigman:2012nb}, a more precise calculation finds a somewhat decreased
value of the cross section needed for the correct thermal relic abundance in 
this mass range, and the required cross section has a significant dependence
on mass.  Second, as mentioned above, these XDM models contain
dark gauge bosons which are mildly relativistic at the time the DM freezes
out; as a result, there are more degrees of freedom in the primordial plasma
than in standard cosmology.  This results in a slight decrease of the 
required cross section, as discussed in \cite{Cline:2010kv}.  Finally,
we assume that the unstable DM excited state has decayed completely by
recombination (see section \ref{decayingconstraints} below), so the average
annihilation cross section changes after chemical freezeout because the
DM states have different relative abundances.  Specifically, at late times,
the stable excited state has relative abundance $\bar Y$ while the ground
state has relative abundance $1-\bar Y$.  Taking into account co-annihilations
between the different DM states, $\langle\sigma_{ann}v_{rel}\rangle$ is
enhanced by a factor $98/75$ at $\bar Y=1/3$, its maximum value, in these
$SU(2)$ triplet models; this enhancement factor increases as $\bar Y$
decreases.  We assume $\bar Y=1/3$.  

Results appear in figure \ref{f:annihilating} for DM mass $M=9$ GeV and
10 GeV.  We see that the exclusion region from the WMAP7+ACT constraint
extends farther to the left (to cuspier halo parameters) than when not 
including DM annihilation (compare to figure \ref{f:xdmconstraints}).  
The improved sensitivity works in both ways:
the WMAP7+ACT constraint rules out these XDM models as thermal WIMPs
for $M<8$ GeV without
XDM-like $e^\pm$ production, but the \textit{Via Lactea II} halo
is excluded for larger masses when both effects are taken into account.
As a further example, the WMAP7+SPT constraint excludes these models 
for $M<18.4$ GeV if XDM-like $e^\pm$ production is ignored but can exclude
the \textit{Via Lactea II} halo for $M<34.5$ GeV when $e^\pm$ production
is included.  At the displayed DM
masses, the annihilation cross sections at chemical freeze out are
$2.39\times 10^{-27}$ cm$^3$/s for $M=$ 9 GeV, $2.34\times 10^{-27}$ cm$^3$/s 
at $M=$ 10 GeV, and $2.10\times 10^{-27}$ cm$^3$/s for $M\geq$ 15 GeV.

\section{Constraints on Decaying Dark Matter}\label{decayingconstraints}

The CMB also constrains energy
deposition to the IGM from decay of DM.  
For lifetimes $\tau$ shorter than
the age of the universe, the exponentially decay of the abundance means
that it is more convenient to constrain $\bar Y_u(\delta M_{12}/M)$, 
$\bar Y_u$ represents the initial abundance of the unstable DM species
relative to all DM.  
Recently,
\cite{Slatyer:2012yq} has provided constraints from WMAP7 data (along with
forecasting Planck constraints), including the dependence of the efficiency
factor $f$ on redshift and the decay product energy $\delta M_{12}$.

It is also important to note that the constraints are irrelevant for decays
with lifetime much shorter than the time of recombination, \textit{ie} 
$\tau\lesssim 10^{13}$ s, since faster decays deposit nearly all of their energy
before the CMB decouples from the IGM.  

\subsection{Decay of Unstable XDM State}\label{xdmdecay}

As noted in \cite{Finkbeiner:2008gw}, XDM models necessarily have at least
one unstable state (which we have denoted as state \#2), and we can ask
what constraints can be placed on the lifetime or initial abundance of
that state.  Here, we update the results of \cite{Finkbeiner:2008gw}.
XDM may also have metastable excited states (denoted state \#3), which must 
have a lifetime longer than the age of the universe, and we will 
consider constraints on this state, as well.

We first consider the metastable state \#3.  For long lifetimes, the WMAP7
constraint becomes $(\delta M_{12}/M)(\bar Y/\tau)<10^{-24.8}$ s$^{-1}$, while
the forecast Planck constraint is $(\delta M_{12}/M)(\bar Y/\tau)<10^{-25.4}$
s$^{-1}$.  Previous studies of XDM indicate that $\bar Y\gtrsim 1/10$
with reasonable assumptions about kinetic freeze out, and it is furthermore
difficult to arrange a sufficient scattering cross section for smaller
relative abundance.  Furthermore, it is reasonable to assume $M\gtrsim 10$ GeV;
with $\delta M_{12}=1.5$ MeV, we find that $\tau\gtrsim 8.9\times 10^{19}$ s
(WMAP7) or $\tau\gtrsim 3.9\times 10^{20}$ s (Planck).

\begin{figure}
\includegraphics[scale=0.6]{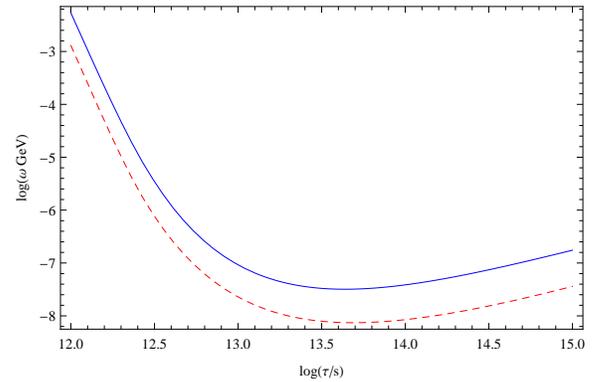}
\caption{\label{f:xdmdecays} Maximum values of 
$\log\omega\equiv\log(\bar Y_u/M)$ (in GeV$^{-1}$) as a function of $\log\tau$ 
(in seconds) from WMAP7 (blue curve) and forecast for Planck (red dashed).
This figure assumes $\delta M_{12}=1.5$ MeV.}
\end{figure}

Turning to the unstable state \#2, \cite{Cline:2012yx} argued that the 
lifetime is constrained by the morphology of the 511 keV signal; if the 
lifetime is greater than approximately $10^{14}$ s, the unstable DM will
move outside the galactic center before emitting the $e^\pm$ pair.
We show constraints from the CMB on $\omega\equiv \bar Y_u/M$ 
for $\delta M_{12}=1.5$ MeV as a function
of lifetime in figure \ref{f:xdmdecays}, using the results of 
\cite{Slatyer:2012yq}.  The effect of the mass splitting $\delta M_{12}$ 
is mainly to shift the constraint curve vertically on the plot, but the
shape of the curve also depends weakly on it.
In most XDM models, $\bar Y_u\sim \bar Y$ due to the
thermal history of DM; in this case, the lifetime can just be as long
as $10^{12}$ s if $M\gtrsim 19$ GeV (WMAP7) or $M\gtrsim 77$ GeV (Planck).
For smaller masses, larger mass splittings, or larger initial relic 
abundances, we simply find $\tau<10^{12}$ s.  Other XDM models, such as
in \cite{Bai:2012yq}, could have much smaller $\bar Y_u$ and therefore
potentially longer lifetimes.  Note that these constraints are potentially
stronger than the constraint from the morphology of the galactic signal
but that the lifetime is degenerate with the mass, mass splitting, and 
relative abundance.  These constraints apply in the case of either
endothermic or exothermic XDM.

We also considered the fact that a long lifetime for the unstable state
could strengthen constraints on downscattering in exothermic XDM models.
Physically, if the unstable state is sufficiently long-lived, $e^\pm$ pairs
are produced at a later cosmological era than the initial downscattering
process.  As a result, the DM density seems to be greater than expected in
$p_{scatt}$.  This process is governed by the Boltzmann equations
\beqa \frac{dn_2}{dt} +3Hn_2&=& -\frac{n_2}{\tau}+n_3^2\langle\sigma_\downarrow 
v_{rel}\rangle\nonumber\\
\frac{dn_3}{dt} +3Hn_3&=& -n_3^2\langle\sigma_\downarrow v_{rel}\rangle\ ,
\label{boltzmann}\eeqa
(ignoring kinetically forbidden upscattering)
and energy deposition is governed by
\beq{longdeposition} 
\frac{dE}{dVdt} = \frac{\delta M_{12} n_2(t)}{\tau}\ .
\eeq
The solutions to the Boltzmann equation for $n_2$ can be given in terms of 
exponential integral functions (in the matter dominated era), 
assuming the abundance of state \#3 to be
constant after early times.  As expected, a lifetime less than the time
of recombination is equivalent to instantaneous decay.  Meanwhile, a longer
lifetime runs into the constraints discussed above unless the initial
abundance of state \#2 is quite small.  In other words, this effect may 
be important in some region of parameter space for models like those of
\cite{Bai:2012yq}.

\subsection{Infalling Positron Model}\label{infallconstraint}

As discussed in section \ref{infall}, \cite{Boubekeur:2012eq} recently
proposed that decaying DM with lifetime in the range $\tau=10^{14}-10^{17}$
s and $\omega\sim 4.7\times 10^{-8}$ GeV$^{-1}$ produces sufficient $e^+$
to explain the galactic 511 keV line.  Note that $\omega$ as defined in
\cite{Boubekeur:2012eq} is our $\bar Y_u$ for $M=100$ GeV; our 
$\omega=\bar Y_u/M$ is a measure of the number density of unstable DM 
particles.\footnote{Another difference is that we use the initial 
relative abundance of 
the unstable DM state, while \cite{Boubekeur:2012eq} uses the value at 
recombination.  Since the lifetimes considered are longer than the time of 
recombination, the difference is unimportant.}  One point of importance
is that the $e^+$ can be produced at high energies due to a mass splitting
of order GeV (or presumably more); they cool quickly in the IGM (mostly
by Compton scattering) before entering the galaxy, where they thermalize
and annihilate.  But it is precisely the cooling process (and annihilation)
in the IGM that modifies the CMB anisotropy spectrum.  While detailed studies 
of the signal morphology for such models have yet to be carried out, we show
here that existing CMB constraints already rule out a great deal of the
parameter space of interest.

\begin{figure}
\includegraphics[scale=0.6]{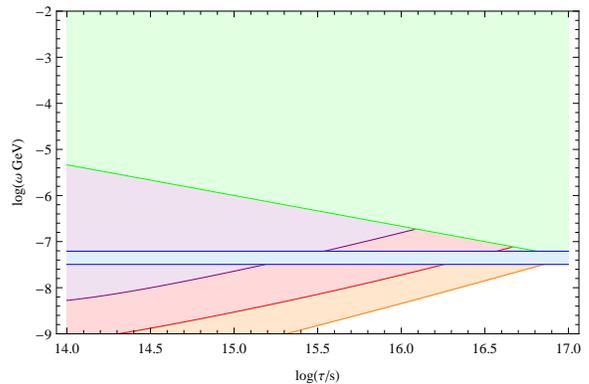}
\caption{\label{f:infalling} Constraints on infalling positron models in the
$log(\tau/$s$)$ and $\log(\omega$ GeV$)$ plane.  The blue band is the 
region of interest for the 511 keV signal.  The top (green) region is excluded
by constraints on diffuse photons for $\delta M_{12}\geq 1$ GeV.  Three
lower regions from bottom to top are excluded by the WMAP7 constraints on
decaying DM for $\delta M_{12}=$ 1 GeV (orange), 100 MeV (red), and 10 MeV
(purple) respectively.}
\end{figure}

Figure \ref{f:infalling} summarizes constraints on these models.  The 
blue band indicates values of $\omega$ identified by 
\cite{Boubekeur:2012eq} as producing the correct number of $e^+$ for the
INTEGRAl signal without overproducing them.\footnote{We have estimated the
width of this band, which is not specified in \cite{Boubekeur:2012eq}.}
The green region is excluded by limits on diffuse photons produced either 
by direct decays or Compton scattering of the resultant $e^+$
as found in \cite{Boubekeur:2012eq}; this 
constraint lifts for mass splittings much below 1 GeV.  The remaining regions
are excluded by the WMAP7 constraints of \cite{Slatyer:2012yq} for
$\delta M_{12}=$ 1 GeV, 100 MeV, and 10 MeV (from bottom to top).  We note
that the preferred value of $\delta M_{12}=$ 1 GeV mentioned in 
\cite{Boubekeur:2012eq} is nearly completely excluded: the central value of 
$\omega=4.7\times 10^{-8}$ GeV$^{-1}$ lies above the WMAP7 constraint for
$\tau\leq 10^{17}$ s and above the diffuse photon constraint for 
$\tau\geq 10^{17}$ s.  Smaller values of the mass splitting $\delta M_{12}$ 
open the parameter space somewhat, but part of the motivation of the
model is lost.  Note that the CMB limits apply equally well to models
in which the unstable DM state is charged and decays by emitting a single
$e^+$ as opposed to an $e^\pm$ pair, since the $e^+$ can still efficiently 
deposit its kinetic energy and annihilate with an ambient $e^-$.  As a result,
we see that existing CMB data provide robust constraints on these models.
Larger mass splittings and the upcoming Planck data will make the constraints
tighter.

\section{Summary}

As we have seen, CMB anisotropies provide robust constraints on DM models
for the galactic bulge $e^+$ production mechanism.  These constraints
provide a new way to exclude annihilating DM and XDM models for a range
of galactic DM halo parameters, including the fiducial parameters from
the \textit{Via Lactea II} simulation and much of the parameter space 
that best fits the 511 keV signal morphology \cite{Vincent:2012an}, 
\textit{as long at the scattering
cross sections are not velocity-suppressed} (or kinematically forbidden) 
at or after recombination and before structure formation.  These models
include $s$-wave annihilation as favored by the signal morphology
\cite{Ascasibar:2005rw} and exothermic XDM models; while the 
\textit{Via Lactea II} parameters are not yet excluded, 
Planck is expected either
to exclude them or find signs of energy deposition.  When including the
effects of DM annihilation, the \textit{Via Lactea II} parameters are 
in fact already ruled out for some XDM models with DM masses less than a 
few tens of GeV, as we illustrated in one case.  

Our most conservative constraints arise because the same scattering 
events that generate the 511 keV gamma ray signal also produce $e^+$
in the early universe, which then deposit their energy in the IGM.
The amount of energy deposition is set by the required rate of $e^+$ 
production in the galactic bulge.  As a result, annihilating light DM 
and exothermic XDM cannot avoid these constraints.  Assuming the 
morphologically-preferred parameter space is ruled out, these DM models
will be viable only if the galactic DM halo profile is cuspier than
usually assumed.  Then the morphology of the 511 keV signal would require
that either the produced $e^+$
propagate significantly before annihilating or that the unstable excited DM
state lives long enough to travel approximately 1 kpc.  Confirmation of
the tentative 130-135 GeV gamma ray line at the galactic center would 
put pressure on this interpretation of the 511 keV signal, however, since
the higher-energy gamma ray morphology appears consistent with the 
\textit{Via Lactea II} halo.  In this case, endothermic XDM could become
the only remaining viable DM model for the 511 keV signal.
A more optimistic point of view, on the other hand, is that
Planck may provide evidence for nonstandard energy deposition in the IGM.
In that case, more detailed studies will be warranted to determine if that
energy deposition is consistent with (or suggestive of)
DM models for the galactic $e^+$ production, as well as potentially 
measuring some parameters of those models.

In this paper, 
we also revisited constraints on the lifetime of the excited DM 
states in XDM models by adapting CMB limits on energy deposition from decaying
DM.  We further used these limits to exclude much of the parameter space of
interest in the recent proposal of \cite{Boubekeur:2012eq} that $e^+$
produced by metastable decaying DM can generate the 511 keV signal when
they fall into the galactic bulge.  

We also note that energy deposition in the SM IGM can also induce
small distortions in the spectrum of the CMB (away from the Planck law)
\cite{Chluba:2011hw,Khatri:2012tv,Khatri:2012tw}.
Similarly, effects of energy deposition should be visible
in the emerging 21 cm window for cosmological observations
\cite{Furlanetto:2006wp,Natarajan:2009bm,Valdes:2012zv}. Either of these
effects may provide interesting constraints on light annihilating DM
or XDM in the future.

In summary,
the ability of DM models to explain the high rate of positron production
in the galactic bulge, along with a striking and suggestive agreement between
anticipated DM halo profiles and the signal morphology, motivates a search for 
additional signals of such DM.  It is clear that CMB anisotropies
provide a new handle on the viability of a broad class of these models
along with their other parameters.

\begin{acknowledgments}
This work was supported in part by the Natural Sciences and Engineering
Research Council (NSERC) of Canada.
\end{acknowledgments}


\end{document}